\documentclass[final,3p,times]{elsarticle}
\usepackage{amsfonts}
\usepackage{mathrsfs}
\usepackage{amsmath}
\usepackage{xcolor}
\usepackage{amssymb}
\usepackage{bm}
\usepackage{feynmp}
\usepackage{extarrows}
\usepackage{slashed}
\usepackage{graphicx}
\usepackage{subfigure}
\usepackage{latexsym}
\numberwithin{equation}{section}
\usepackage[linktocpage=true,bookmarks=true,bookmarksnumbered=true,breaklinks=true,pdfpagemode=Fullscreen,pdfstartview=FitBH]{hyperref}

\newcommand{\fr}[2]{\mbox{$\frac{\,{#1}\,}{#2}$}}

\def\bge{\begin{equation}}
\def\ede{\end{equation}}
\def\bga{\begin{aligned}}
\def\eda{\end{aligned}}
\newcommand{\beq}{\begin{equation}}
\newcommand{\eeq}{\end{equation}}
\newcommand{\bq}{\begin{equation}}
\newcommand{\eq}{\end{equation}}
\newcommand{\ba}{\begin{array}}
\newcommand{\ea}{\end{array}}
\newcommand{\beqa}{\begin{eqnarray}}
\newcommand{\eeqa}{\end{eqnarray}}
\newcommand{\beqs}{\begin{subequations}}
\newcommand{\eeqs}{\end{subequations}}

\def\dis{\displaystyle}

\def\({\left(}
\def\){\right)}
\def\[{\left[}
\def\]{\right]}

\def\End{\end{document}}

\def\ii{\mathrm{i}}

\def\over{\overline}

\def\leqq{\leqslant}
\def\geqq{\geqslant}

\def\hf{\frac{1}{2}}

\def\X{\chi}

\def\lam{\lambda}

\def\si{\sigma}

\def\X{\chi}
\def\et{\eta}
\def\vet{v_\eta^{}}

\def\vphi{v_\phi^{}}

\def\dd{\mathrm{d}}

\def\B{\mathcal{B}}

\def\to{\rightarrow}

\def\ii{\mathrm{i}}

\def\bbbar{b\bar{b}}
\def\tautaub{\tau\bar{\tau}}

\def\N{\mathcal{N}}
\def\A{\mathbb{A}}
\def\B{\mathbb{B}}

\def\cut{\Lambda_{\textrm{UV}}^{}}

\def\End{\end{document}}
\newcommand{\pbrac}[1]{\left( #1 \right)}
\newcommand{\tbrac}[1]{\left[ #1 \right]}
\newcommand{\cbrac}[1]{\left\{ #1 \right\}}

\begin{document}

\begin{frontmatter}

\setcounter{footnote}{1}
\renewcommand{\thefootnote}{\fnsymbol{footnote}}

\title{{\bf Natural Electroweak Symmetry Breaking \\
            from Scale Invariant Higgs Mechanism}}

\author{Arsham Farzinnia$^{\,1}$\footnote{farzinnia@tsinghua.edu.cn},~~
        Hong-Jian He$^{\,1,2,3}$\footnote{hjhe@tsinghua.edu.cn},~~
        Jing Ren$^{\,1}$\footnote{renj08@mails.tsinghua.edu.cn}%
        }

\address{$^1$\,Institute of Modern Physics and Center for High Energy Physics,
               Tsinghua University, Beijing 100084, China\\
         $^2$\,Center for High Energy Physics, Peking University, Beijing 100871, China\\
         $^3$\,Kavli Institute for Theoretical Physics China, CAS, Beijing 100190, China}

\begin{abstract}
We construct a minimal viable extension of the standard model (SM) with
classical scale symmetry. Its scalar sector contains a complex singlet
in addition to the SM Higgs doublet. The scale-invariant and CP-symmetric
Higgs potential generates radiative electroweak symmetry breaking
\`{a} la Coleman-Weinberg, and gives a natural solution to the hierarchy problem,
free from fine-tuning. Besides the 125\,GeV SM-like Higgs particle, it predicts
a new CP-even Higgs (serving as the pseudo-Nambu-Goldstone boson of
scale symmetry breaking) and a CP-odd scalar singlet (providing
the dark matter candidate) at weak scale.
We systematically analyze experimental constraints from direct
LHC Higgs searches and electroweak precision tests, as well as
theoretical bounds from unitarity, triviality and vacuum stability.
We demonstrate the viable parameter space, and discuss implications for
new Higgs and dark matter (DM) searches at the upcoming LHC runs and
for the on-going direct detections of DM.
\\[2mm]
\\[2mm]
\hfill
Phys.\ Lett.\ B (2013), in Press [arXiv:1308.0295]
\end{abstract}


\end{frontmatter}




\setcounter{footnote}{0}
\renewcommand{\thefootnote}{\arabic{footnote}}

\section{Introduction}
\vspace*{2mm}
\label{intro}

The LHC discovery of a 125\,GeV Higgs-like particle \cite{LHCnew}\cite{LP2013} seems to
provide the last missing piece of the standard model (SM) of particle physics,
but the SM apparently fails to accommodate dark matter (DM) and neutrino masses.
Higgs mechanism \cite{HM} is the cornerstone of the SM,
which hypothesizes a single spin-0 Higgs doublet
to realize the spontaneous electroweak symmetry breaking (EWSB) and gives rise to
a physical remnant --- the Higgs boson.
This generates \cite{weinberg} the observed masses for spin-1 weak bosons and
all three families of spin-$\frac{1}{2}$ SM fermions via
gauge and Yukawa interactions of the Higgs boson.
However, the Higgs boson could not fix its own mass and an {\it ad hoc}
negative mass term is input by hand at the weak scale.
As such, it is customary to think that the Higgs mass will be unstablized
against the Planck scale by quantum corrections
unless large fine-tuned cancellation
of the associated quadratical divergences is imposed \cite{fine-tuning}.
Historically, seeking resolutions to this naturalness problem
has been the major driving force behind numerous ``beyond SM" extensions on the market,
ranging from supersymmetry and compositeness to large or small extra dimensions,
despite none of them has been seen so far at the LHC.

The naturalness theorem \cite{natural} asserts that the absence of large
corrections can only be maintained through certain symmetry which protects the
Higgs mass term. This means that the symmetry must increase when the Higgs
mass approaches zero. It is important to note that the Higgs mass is the unique dimensionful
parameter in the SM Lagrangian, and only causes soft breaking of the
scale symmetry.\footnote{After the SM is extended with singlet right-handed neutrinos,
their dimension-3 heavy Majorana mass-term provides another soft breaking of scale invariance.
Our present construction will naturally generate this Majorana mass term via spontaneous
symmetry breaking.}\,
Such a scale symmetry will also be explicitly broken
by the trace anomaly with dimension-4 operators at quantum level.
But this only leads to logarithmic running of coupling constants
and cannot generate quadratical divergence in the dimension-2 Higgs mass term \cite{Bardeen}.
Hence, the SM itself could be technically natural up to high scales\footnote{The SM Higgs
sector with a 125\,GeV Higgs boson is free from triviality bound, but suffers
a vacuum stability bound at the scale $\,\mu\simeq 10^{12}\,$GeV \cite{Espinosa2012}.
We will analyze both triviality and vacuum stability bounds for the present model.}
and free from fine-tuning in the Higgs mass renormalization
due to the softly broken classical scale invariance \cite{Bardeen}\cite{xtalk}.

It is even more tempting to restore the full scale symmetry for the SM Lagrangian
by setting a vanishing Higgs mass.  This justifies the use of a scale-invariant
regularization method for loop corrections, which automatically ensures the absence
of quadratical divergence in the Higgs mass renormalization. (The simplest regulator
respecting classical scale symmetry is the dimensional regularization \cite{DR}.)
Thus, such a scale-invariant SM Lagrangian or its scale-symmetric extensions
will stabilize the weak scale up to a high ultraviolet (UV) cutoff $\,\cut\,$
provided \cite{Bardeen}:
{\it (1).}\ no intermediate scales\,\footnote{Our present model
will extend the scale-invariant SM Lagrangian
with a complex Higgs singlet and three right handed neutrinos at TeV scale.
Hence, it is a technically natural {\it effective field theory} (EFT),
all the way up to its UV cutoff (above which a more complete theory
arises and is assumed to properly retain classical scale symmetry).
We do not concern detailed Planck-scale dynamics \cite{UV}, given the lack
of a full theory of quantum gravity.
This EFT is also free from little hierarchy problem because it invokes no extra heavy state
at intermediate scales. We thank Nima Arkani-Hamed for discussing this point.}
would mix with the weak scale;\,
{\it (2).}\ no Landau poles (or instabilities) appear in the running couplings
(or Higgs potential) over the energies up to $\,\cut\,$.\,

With such a fully scale-invariant SM Lagrangian, one can radiatively generate
nonzero Higgs mass and spontaneous EWSB via Coleman-Weinberg mechanism \cite{CW}.
In consequence, the weak scale is nicely induced at quantum level via dimensional transmutation.
This further reduces one more free-parameter from the conventional SM.
But, unfortunately such a minimal version has its Higgs potential unbounded from below
at one-loop given the experimentally observed masses of top quark and weak gauge bosons.
In addition, the radiatively induced Higgs mass is too low to even survive
the LEP-II Higgs search bound  $\,M_h > 114.4\,$GeV (95\%\,C.L.) \cite{LEP2-mH}.
Hence, the SM Higgs sector has to be properly extended and some interesting attempts appeared in
recent years \cite{SI-other,SI-long}.

In this work, we construct the minimal viable extension of the SM with classical scale symmetry.
Its Higgs sector contains a Higgs doublet and a complex gauge-singlet scalar.
The Higgs potential is scale-invariant, as well as CP-conserving.
The model predicts two CP-even Higgs boson and one CP-odd scalar at weak scale.
Among the two CP-even states, one provides the observed 125\,GeV Higgs boson and another
serves as a pseudo-Nambu-Goldstone boson from scale symmetry breaking.
The CP-odd scalar is a potential dark matter candidate.
We will demonstrate that including the complex singlet scalar
not only helps to lift the radiative mass of the Higgs boson to coincide with
the current LHC Higgs data \cite{LHCnew}\cite{LP2013},
but also nicely generate the Majorana mass term for right-handed neutrinos
from scale-invariant Yukawa interaction.
We systematically analyze experimental and theoretical constraints
on the parameter space of our model.
These include experimental bounds from the direct LHC Higgs measurements
and the indirect electroweak precision tests, as well as the
theoretical constraints from unitarity, triviality and vacuum stability.
Finally, we note that our approach also differs from
the previous studies \cite{SI-other,SI-long} (\`{a} la Coleman-Weinberg)
invoking extra scalars or certain hidden gauge groups.
Those extended gauge groups include the $U(1)_{X}^{}$
(sometimes $U(1)_{B-L}^{}$), or the left-right gauge group,
or the vector dark $SU(2)_D^{}$, or certain strongly interacting hidden sector.
An extensive analysis of a complex singlet scalar
with the global $U(1)$ (or $Z_4$) symmetry
and maximal CP-violation was given in \cite{SI-long},
which differs from our CP-symmetric and scale-invariant Higgs sector
(without extra global or local symmetry).
Our model also differs from \cite{JS2-Z2} which considered two real
scalar singlets with an extra $Z_2$ to ensure stability of the $Z_2$-odd
singlet as DM. In contrast, our model builds the imaginary component of the
complex singlet as DM and its stability is automatically protected by CP invariance;
we further include right-handed neutrinos for light neutrino mass-generations
via TeV scale seesaw.

This paper is organized as follows. Sec.\,\ref{model} sets up
the model construction for our classically scale-invariant Higgs potential.
Then, we present the full one-loop corrections,
identify the physical states, and derive their mass spectrum and couplings.
In Sec.\,\ref{bounds} we study both experimental and theoretical constraints
on the parameter space of the model. Sec.\,\ref{results} presents our results
and discusses the physical implications.
Finally, we conclude in Sec.\,\ref{conclusion}.


\vspace*{2mm}
\section{Model Structure and Radiative Electroweak Symmetry Breaking}
\label{model}

In this section, we construct the minimal viable extension of the SM with
classical scale-invariance. It only contains an extra gauge-singlet complex
scalar \,$S$\, in addition to the conventional Higgs doublet $\,H\,$.\,
Our extended Higgs sector is CP invariant (similar to the SM) and
respects the classical scale symmetry. This will naturally induce radiative
EWSB and predict two new scalar states in addition to the observed 125\,GeV
light Higgs boson.  This minimal construction maximally preserves all the
original SM symmetries, and further incorporates three right-handed neutrinos
for mass-generation of light neutrinos via TeV scale seesaw.

\vspace*{2mm}
\subsection{{\bf The Model Structure}}
\label{Sec:2.1}
\vspace*{2mm}

In our construction, the extended Higgs sector consists of the SM Higgs doublet $\,H\,$
and a complex singlet scalar $\,S$\,,\, so its Lagrangian is,
\begin{equation}
\label{Lscalar}
\mathcal{L}_{S}^{} ~=~
(D^\mu H)^\dagger D_\mu^{} H +
\partial^\mu S^* \partial_\mu S - V^{(0)}(H,S) \ ,
\end{equation}
where the Higgs doublet $\,H$\, is expressed in component form,
\begin{equation}
\label{H}
H ~=\,
\begin{pmatrix}
\pi^+
\\[1.5mm]
\fr{1}{\sqrt{2}\,}\(v_\phi^{}+\phi^0+\ii \,\pi^0\)
\end{pmatrix} ,
\end{equation}
and $\,D^\mu\,$ is the covariant derivative under SM gauge group.
In \eqref{H}, $\,\phi\,$ is the SM-like Higgs field, with the vacuum expectation value (VEV),
\,$v_\phi^{} \simeq 246$\,GeV, to be determined from radiative EWSB.
The gauge-singlet scalar field $\,S\,$ has the following component form,
\begin{equation}
\label{S}
S ~=\, \fr{1}{\sqrt{2}\,} \(\,v_\eta^{} + \eta^0_{} + \ii\,\chi^0_{}  \,\) ,
\end{equation}
where $\,\et\,$ has $\,J^P=0^+$.\, Thus, under either \,C\, or \,CP\, operation it transforms as,
$\,S\to S^*\,$.\, This means that $\,\eta\,$ and $\,\X\,$
belong to the CP-even and CP-odd fields, respectively.

Then, we can write down the most general scale-invariant and CP-symmetric Higgs potential
with the Higgs doublet $\,H\,$ and complex singlet $\,S\,$,
\begin{equation}
\label{V0}
V^{(0)}_{}(H,S)~ =\, \frac{\lambda_1^{}}{6} \pbrac{H^\dagger H}^2 +
\frac{\lambda_2^{}}{6} |S|^4 +
\lambda_3^{} \pbrac{H^\dagger H}|S|^2 +
\frac{\lambda_4^{}}{2} \pbrac{H^\dagger H}\pbrac{S^2 \!+ S^{*2}} +
\frac{\lambda_5^{}}{12} \pbrac{S^2 \!+ S^{*2}} |S|^2 +
\frac{\lambda_6^{}}{12} \pbrac{S^4 \!+ S^{*4}} \, ,~~~
\end{equation}
which contains six dimensionless real coupling constants $\,\{\lambda_j^{}\}$\,.\,
Here the cubic couplings and mass terms are forbidden by the scale-invariance.
In the above potential,
the general mixing between Higgs doublet and singlet is represented by the third
and fourth terms via the quartic couplings
$\,\lambda_3^{}\,$ and $\,\lambda_4^{}\,$.\,
For practical analysis, we find it convenient to introduce
the following coupling combinations,\footnote{The coupling normalizations in \eqref{V0} and
\eqref{couprel} have been chosen such that the associated Feynman
rules of the scalar quartic interactions take the simple form of
$\,\pm \ii\,\lambda_j^{}$\,.}
\begin{equation}
\label{couprel}
\lambda_\phi^{} \equiv \lambda_{1}^{}\,, \quad
\lambda_\et^{} \equiv \lambda_{2}^{} + \lambda_{5}^{} + \lambda_{6}^{}\,, \quad
\lambda_\X^{} \equiv \lambda_{2}^{} - \lambda_{5}^{} + \lambda_{6}^{}\,, \quad
\lambda_{\et\X}^{} \equiv \fr{1}{3}\lambda_{2}^{} - \lambda_{6}^{} \,, \quad
\lambda_m^+  \equiv \lambda_{3}^{} + \lambda_{4}^{} \,, \quad
\lambda_m^-  \equiv \lambda_{3}^{} - \lambda_{4}^{} \,.~~
\end{equation}
Thus, we infer the quartic scalar interactions in terms of component fields,
\begin{equation}
\label{V0quart}
\begin{split}
V^{(0)}_{} ~=~&
\fr{1}{24} \tbrac{ \lambda_\phi^{} \phi^4 +
\lambda_\et^{} \et^4 + \lambda_\X^{} \X^4 +
\lambda_\phi^{}\pbrac{\pi^0\pi^0 + 2 \pi^+ \pi^-}^2}+
\fr{1}{4}\(\lambda_m^+ \phi^2 \et^2 + \lambda_m^- \phi^2\X^2 +\lambda_{\et\X}^{}\et^2\X^2\)
\\[2mm]
& + \fr{1}{12} \(\lambda_{\phi}^{} \phi^2 + 3\lambda_m^+ \et^2 +
    3\lambda_m^- \X^2\)\pbrac{\pi^0\pi^0 + 2 \pi^+ \pi^-} \,.
\end{split}
\end{equation}
In terms of these variables, the tree-level potential \eqref{V0} or \eqref{V0quart} is
bounded from below under the conditions,
\beqs
\label{eq:Vstability-tree}
\begin{align}
&
\lambda_\phi^{} > 0 \ , \qquad \lambda_\et^{} > 0 \ , \qquad
\lambda_\X^{} > 0 \,, \qquad
\pbrac{\lambda_m^+}^2 < \fr{1}{9} \lambda_\phi^{} \lambda_\et^{} \,, \qquad
\pbrac{\lambda_m^-}^2 < \fr{1}{9} \lambda_\phi^{} \lambda_\X^{} \,,
\label{stabtree1}
\\[2mm]
&
\lambda_{\et\X}^{} > -\fr{1}{3}\!\sqrt{\lambda_{\et}^{}\lambda_{\X}^{}} \,, \qquad
\lambda_\phi^{} \lambda_{\et\X}^{} - 3 \lambda_m^+ \lambda_m^-
> -\fr{1}{3}\!
\sqrt{\tbrac{\lambda_\phi^{}\lambda_\et^{} \!-\! 9\pbrac{\lambda_m^+}^2}
\tbrac{\lambda_\phi^{} \lambda_\X^{} \!-\! 9\pbrac{\lambda_m^-}^2}\,} \,.
\qquad
\label{stabtree3}
\end{align}
\eeqs

Finally, we include three right-handed Majorana neutrinos, which will
account for the observed light neutrino masses via seesaw mechanism \cite{seesaw}.
In our construction, we conjecture that the pure singlet sector
(including singlet scalar $S$ and singlet neutrino $\N$) always conserves CP\,.\,
This requires the Yukawa interactions between $\,S\,$ and
$\,\N\,$ to be CP symmetric.
Thus, we can write down the gauge- and scale-invariant Yukawa interactions
for neutrino sector,
\begin{equation}
\label{LRHN}
\mathcal{L}_{\nu}^{} ~=\,
- \(Y^\nu_{ij} \bar{L}_{iL}^{} \widetilde{H} \N_j^{} + \text{h.c.}\)
-\fr{1}{2}Y^N_{ij}\(S + S^*\) \overline{\N}_i^{}\N_j^{} \,,
\end{equation}
where  $\,\widetilde{H} = \ii\,\sigma^{}_2 H^*$,\, and
$\,\N_j^{} = \N_j^c\,$ is a 4-component Majorana spinor
denoting the singlet (right-handed) neutrinos.
Our construction builds the singlet neutrino $\,\N_j^{}\,$
as a Majorana spinor starting from the symmetric phase, and
$\,\N_j^{}\,$ will acquire Majorana mass after spontaneous scale symmetry breaking.
In the above, $\,\{Y^\nu_{ij}\}\,$ denotes Yukawa couplings of the
Higgs doublet $\,H\,$ with left-handed lepton doublet $\,L_{L}^{i}\,$
and singlet neutrino $\,\N_j^{}\,$, while $\,Y^N\, $
represents the Yukawa couplings between the singlet Higgs $\,S\,$ and
singlet Majorana neutrinos $\,\N\,$.\,
It is straightforward to verify CP invariance of the above
$\,S$-$\N$-$\over{\N}\,$ Yukawa intecations since $\(S + S^*\)$ and
$\,\overline{\N}_i^{}\N_j^{}\,$ respect CP symmetry, respectively.
Besides, since the operator $\(S + S^*\)\overline{\N}_i^{}\N_j^{}\,$
equals its own Hermitian conjugate, the Yukawa couplings $\,Y^N_{ij}\,$ are real.
In the practical analysis, we will always choose $\,Y^N\,$ in the diagonal basis,
and for simplicity we set $\,Y^N\,$ to be degenerate,
$\, Y^N = y_N^{} {\cal I}_{3\times 3}^{}$.\,
We note that because our gauge-singlet sector conserves CP,
the CP-odd scalar $\,\X\propto \mathrm{Im}(S)\,$ has vanishing Yukawa coupling with
the singlet neutrinos $\,\N\,$ in Eq.\,(\ref{LRHN}).
This is a key feature of our model which ensures that the pseudo-scalar $\,\X\,$
always appears in pair via CP-invariant Higgs potential \eqref{V0quart} and is thus stable.
Hence, the $\,\X\,$ boson provides a natural dark matter candidate.

We note that the Yukawa interactions (\ref{LRHN}) will generate seesaw masses
for light neutrinos,
\beqa
\label{eq:seesaw}
m_\nu^{} ~=~ m_D^{}M_N^{-1}m_D^T \,,
\eeqa
where $\,m_D^{}= Y^\nu v_\phi^{}/\sqrt{2}\,$  and
$\,M_N=\sqrt{2}Y^Nv_\et^{}\,$.\,
For our construction, we will set the singlet scalar VEV
$\,v_\et^{}=O(\text{TeV})\,$.\,
Inputting the scale of light neutrino masses from oscillation data
$\,m_\nu^{}= O(0.1\text{eV})\,$ and taking
$\,Y^N=O(1)\,$,\, we find,
$\,m_D^{}=O(m_e^{})\,$ with $\,m_e^{}\,$ the electron mass.
Thus, $\,Y^\nu = O(m_e^{}/v_\phi^{})\,$  is around the size of
the electron Yukawa coupling of the SM. Hence, in the following effective potential analysis
we can safely ignore the tiny Dirac Yukawa coupling $\,Y^\nu$,\,
and only retain the Majorana Yukawa term in (\ref{LRHN}).

\vspace*{2mm}
\subsection{{\bf Mass Eigenvalues at Tree-Level}}
\label{Sec:2.2}
\vspace*{2mm}

For the present study, we will determine the physical vacuum and Higgs mass-eigenvalues
from minimizing the full scalar potential up to one-loop,
\begin{equation}\label{V01}
V(H,S) ~=~ V^{(0)}_{}(H,S) + V^{(1)}_{}(H,S) \ ,
\end{equation}
where $\,V^{(1)}_{}(H,S)$\, is the one-loop contribution
from all relevant fields running in the loop.
Such a minimization with multiple scalars is technically complicated
in general.  Following the approach of Gildener and Weinberg \cite{Gildener:1976ih},
we first minimize the tree-level potential \eqref{V0}, and
keep in mind that the potential couplings become running
at quantum level and depend on the renormalization scale $\,\mu$\,.\,
Thus, the minimization of the tree-level potential is performed at a particular scale
$\,\mu = \Lambda$\,,\, and gives a ``flat'' direction among the scalar VEVs.
Further including one-loop corrections will lift this flat direction and generate
the true physical vacuum (corresponding to the spontaneous breaking of
classical scale invariance).

Starting from the tree-level scalar potential \eqref{V0}, we analyze its minimization
with respect to the Higgs fields $H$ and $S$ at the scale $\,\mu = \Lambda\,$,
and derive the conditions,
\begin{equation}
\label{mincond}
\dis
\frac{\,v_\phi^2\,}{v_\et^2} \,=\,
\frac{\,-3\lambda_m^+(\Lambda)\,}{\lambda_\phi^{}(\Lambda)}
\,=\, \frac{\lambda_\et^{}(\Lambda)}{\,-3\lambda_m^+(\Lambda)\,} \ .
\end{equation}
This defines the flat direction of the potential, and further
implies [cf.\ \eqref{stabtree1}],
$\,\lambda_\phi^{}(\Lambda)>0$,\,
$\lambda_\et^{}(\Lambda)>0$,\, and
\beqa
\label{couplflat}
\lambda_m^+(\Lambda) ~=\,
-\fr{1}{3}\!\sqrt{\lambda_\phi^{}(\Lambda)\lambda_\et^{}(\Lambda)\,} \,,
&~~\text{or,}~~&
\lambda_\et^{}(\Lambda)
~=~\frac{\,9\lambda_m^+(\Lambda)^2\,}{\lambda_\phi^{}(\Lambda)} \,.
\eeqa

Then, we can compute the tree-level mass spectrum from the scalar potential \eqref{V0}.
Expanding the Higgs fields in terms of their components \eqref{H}-\eqref{S},
and using the definitions \eqref{couprel}, we deduce the quadratic Higgs mass-terms,
\begin{equation}
\label{V0quad}
\begin{split}
V^{(0)}_{\text{mass}} ~=~ &\,
\fr{1}{4}
\begin{pmatrix} \phi \!&\! \et \end{pmatrix}
\begin{pmatrix}
\lambda_\phi^{} v_\phi^2+\lambda_m^+ v_\et^2 & 2\lambda_m^+ v_\phi^{} v_\et^{}
\\[2mm]
2\lambda_m^+ v_\phi^{} v_\et^{} & \lambda_\et^{} v_\et^2 + \lambda_m^+ v_\phi^2
\end{pmatrix}
\begin{pmatrix} \,\phi\, \\[2mm] \,\et\, \end{pmatrix}
\\[2mm]
&\, + \fr{1}{4} \pbrac{\lambda_m^- v_\phi^2 + \lambda_{\et\X}^{} v_\et^2}\X^2 +
\fr{1}{12}\pbrac{\lambda_{\phi}^{} v_\phi^2 + 3\lambda_m^+ v_\et^2}
\pbrac{\pi^0\pi^0 + 2 \pi^+ \pi^-} \,.
\end{split}
\end{equation}
The mass terms of the CP-even components \,$(\phi,\,\eta )$\, form a $2\times 2$ matrix,
and can be diagonalized by an orthogonal rotation,
\begin{equation}
\label{hs}
\begin{pmatrix} \,\phi\, \\[1mm] \,\et\, \end{pmatrix}
~=~ \mathbb{O}
\begin{pmatrix} \,h\, \\[1mm] \,\sigma\, \end{pmatrix} \,,
\end{equation}
where $\,(h,\,\sigma)$ are the CP-even mass-eigenstates.
The rotation matrix $\,\mathbb{O}\,$ is defined with mixing
angle $\,\omega\,$,\, and can be determined as follows,
\beqa
\label{O}
\mathbb{O} \,\equiv\,
\begin{pmatrix} \cos\omega & \sin\omega
\\[2mm]
-\sin\omega & \cos\omega \end{pmatrix} \,, \qquad
\cot 2\omega \,\equiv\,
\frac{1}{4\lambda_m^+}
\tbrac{ \(\lambda_\et^{}-\lambda_m^+\)
\frac{v_\et^{}}{v_\phi^{}} - \(\lambda_\phi^{}-\lambda_m^+\) \frac{v_\phi^{}}{v_\et^{}}} .
\eeqa
Accordingly, we derive the tree-level mass-eigenvalues for all scalar states
after mass-diagonalization,
\beq
\ba{ll}
\label{masstree}
\dis M_h^2 ~=\,
\fr{1}{2}
\pbrac{\lambda_\phi^{} v_\phi^2+\lambda_m^+ v_\et^2
       - 2\lambda_m^+ v_\phi^{} v_\et^{} \tan\omega} ,
\quad~~
&
\dis M_\sigma^2 ~= \fr{1}{2}
\pbrac{\lambda_\phi^{} v_\phi^2+\lambda_m^+ v_\et^2 + 2\lambda_m^+ v_\phi^{}v_\et^{} \cot\omega} ,
\\[3mm]
\dis M_\X^2 ~=\, \fr{1}{2} \pbrac{\lambda_m^- v_\phi^2 + \lambda_{\et\X}^{} v_\et^2} ,
&
\dis M_{\pi^0}^2 ~=\, M_{\pi^\pm}^2 \,=\,
\fr{1}{6}\pbrac{\lambda_{\phi}^{}v_\phi^2 + 3\lambda_m^+ v_\et^2} .
\ea
\eeq
Besides, from Eq.\,\eqref{LRHN} and
taking $\,Y^N =y_N^{}{\cal I}_3^{}\,$ for simplicity,
we infer a tree-level mass formula for right-handed neutrinos,
\begin{equation}
\label{mN}
M_N^{} ~=~ \sqrt{2} \,y_N^{} v_\et^{} \ .
\end{equation}
Implementing the minimum condition \eqref{mincond} at
scale $\,\mu = \Lambda\,$,\,  we obtain,
$\,\cot\omega = {v_\et^{}}/{v_\phi^{}}$\,
for the mixing angle, and reexpress the tree-level masses,
\begin{equation}
\label{masstreemincond}
\ba{ll}
\dis M_h^2 ~=~ \frac{v_\phi^2}{3} \tbrac{\lambda_\phi^{}(\Lambda) -3\lambda_m^+(\Lambda) } \,,
\quad
&
\dis M_\X^2 ~=~  \frac{v_\phi^2}{\,6\lambda_m^+(\Lambda)\,}
\tbrac{3\lambda_m^+(\Lambda)\lambda_m^-(\Lambda)
       - \lambda_{\phi}^{}(\Lambda)\lambda_{\et\X}^{}(\Lambda)} \,,
\\[4mm]
\dis
M_\sigma^2 ~=~ M_{\pi^0}^2 ~=~ M_{\pi^\pm}^2 ~=~ 0 \,,
\quad
&
\dis M_N^{} ~=~ y_N^{} v_\phi^{}
\sqrt{\frac{2\lambda_\phi^{}(\Lambda)}{\,-3\lambda_m^+(\Lambda)\,}\,}
\ .
\ea
\end{equation}
As expected, we find three massless would-be Nambu-Goldstone bosons $\,\{\pi^\pm_{},\,\pi^0_{}\}\,$
eaten by $\,\{W^\pm,\,Z^0\}\,$,\, and another massless
CP-even state $\,\sigma$\, which is the Nambu-Goldstone boson of
spontaneously broken classic scale symmetry.
As will be shown below, the $\,\sigma$\, particle will acquire its radiative mass
along the flat direction \`{a} la Coleman-Weinberg \cite{CW}, and
thus becomes a pesudo-Nambu-Goldstone boson at quantum level.
Hence, we have only two massive scalar bosons at tree-level,
the CP-even state $\,h\,$ and the CP-odd component $\,\X\,$.\footnote{We
note that it is possible to implement an inverse identification
of the CP-even states, such that $\,h\,$ becomes a tree-level massless
state (and thus the pseudo-Nambu-Goldstone boson of scale symmetry breaking),
whereas $\,\sigma\,$  acquires nonzero mass at tree-level.
But, as we will find in Sec.\,\ref{results}, this setup is excluded by
the theoretical and experimental constraints.}

We note that the pseudo-scalar $\,\X\,$ is protected by CP symmetry,
namely, because of the CP conservation associated with the Higgs potential \eqref{V0}
and singlet Yukawa sector (involving $S$ and $\N$),
the $\,\X\,$ boson always appears in pair and thus provides a stable dark matter candidate.
In the following analysis, we will identify the CP-even Higgs boson $\,h\,$ with the
125\,GeV new state discovered at the LHC \cite{LHCnew}\cite{LP2013}.
In Sec.\,\ref{results}, we will find that the other way of identifying $\,\si\,$ boson with
the 125\,GeV state is excluded by the current data.

\vspace*{2mm}
\subsection{{\bf Radiative EWSB from One-Loop Effective Potential}}
\label{Sec:2.3}
\vspace*{2mm}

So far we have been working on the tree-level Higgs potential \eqref{V0},
where the flat direction \eqref{mincond} does not pick up any true physical vacuum
for the EWSB.  Therefore, it is important to compute the one-loop potential
$\,V^{(1)}_{}$.\,
This will become dominant along the flat direction \eqref{mincond}, and
thus produce realistic radiative EWSB \`{a} la Coleman-Weinberg.
According to E.\ Gildener and S.\ Weinberg \cite{Gildener:1976ih},
we cast the one-loop effective potential
into the general form at the renormalization scale $\,\mu = \Lambda$\,,
\beqa
\label{V1}
V^{(1)}(\varphi) ~=~ \mathbb{A}\,\varphi^4 + \mathbb{B}\,\varphi^4 \log
\frac{\varphi^2}{\Lambda^2} \ ,
\eeqa
where $\,\varphi\,$ is the ``radial" combination of the Higgs fields at
$\,\mu = \Lambda$\,,
\beqa
\label{phi}
\varphi^2 ~=~ \phi^2(\Lambda) + \et^2(\Lambda) ~=~
\frac{\,\phi^2(\Lambda)\,}{\,\sin^2 \!\omega\,} \,.
\eeqa
Since the one-loop potential (\ref{V1}) is computed at $\,\mu = \Lambda$\,
and along the flat direction \eqref{mincond},
the relation $\,\cot\omega=v_\eta/v_\phi = \eta(\Lambda)/\phi(\Lambda)\,$ holds,
as inferred below \eqref{mN}.  From this we can deduce the second equality of \eqref{phi}.
The coefficients $\,\mathbb{A}\,$ and $\,\mathbb{B}\,$ are dimensionless loop-generated
constants, under the $\overline{\text{MS}}$ scheme
\cite{Gildener:1976ih,SI-long},
\beqs
\label{AB}
\beqa
\mathbb{A} &\!\!=\!\!\!& \frac{1}{\,64\pi^2 v_\varphi^4\,} \cbrac{\text{Tr}\tbrac{M_S^4\pbrac{-\frac{3}{2}+\log\frac{M_S^2}{v_\varphi^2}}} + 3\text{Tr}\tbrac{M_V^4\pbrac{-\frac{5}{6}+\log\frac{M_V^2}{v_\varphi^2}}} - 4\text{Tr}\tbrac{M_F^4\pbrac{-1+\log\frac{M_F^2}{v_\varphi^2}}}} \,,
\hspace*{12mm}
\\[1mm]
\mathbb{B} &\!\!=\!\!\!& \frac{1}{\,64\pi^2 v_\varphi^4\,} \pbrac{\text{Tr}\,
M_S^4 + 3\text{Tr}\, M_V^4 - 4\text{Tr}\, M_F^4} \, ,
\eeqa
\eeqs
where the traces are taken over all internal degrees of freedom,
and $\,M_{V,S,F}^{}$\, represent involved tree-level masses of vectors, scalars
and fermions evaluated at $\,\mu=\Lambda$\,.\,
This scale may be determined from minimizing the one-loop
potential \eqref{V1},
\,${\dd V^{(1)}(\varphi)}/{\dd\varphi}\big|_{\varphi = v_{\varphi}}^{} = 0$\,,\,
yielding
\begin{equation}
\label{Lambda}
\Lambda ~=~ v_\varphi^{} \exp\tbrac{\frac{\A}{\,2\B\,}+\frac{1}{4}} \ .
\end{equation}
Moreover, the one-loop potential $\,V^{(1)}_{}\,$
will generate a mass term for the $\,\sigma\,$
pseudo-Nambu-Goldstone boson along the flat direction.
With \eqref{Lambda}, we compute this loop-induced $\,\sigma\,$ mass,
\begin{equation}
\label{ms}
M_\sigma^2 ~=~ \left.\frac{\dd^2V^{(1)}(\varphi)}{\dd\,\varphi^2}
\right|_{\varphi = v_{\varphi}^{}} =~ 8v_{\varphi}^2 \B \ .
\end{equation}

For the present model, we consider the relevant tree-level masses,
$M_h^{}$, $M_\X^{}$, $M_W^{}$, $M_Z^{}$, $M_t^{}$, and $M_N^{}$, which include
the masses of scalars and right-handed neutrinos in
\eqref{masstreemincond}, as well as the heavy SM fields of top quark and
$(W,\,Z)$ vector bosons. With the aid of \eqref{mincond} and
\eqref{phi}, we can write down the one-loop potential \eqref{V1}
in terms of $\,\phi\,$ and its VEV, $\,\vphi \simeq 246$\,GeV,
\beqa
\label{V1h0}
V^{(1)}_{}(\phi) ~=~ \A^\prime \phi^4 + \B^\prime \phi^4 \log \frac{\phi^2}{\Lambda^2}
\ ,
\eeqa
with the coefficients under $\overline{\text{MS}}$ scheme,
\beqs
\label{ApBp}
\beqa
\label{eq:A'}
\A^\prime &\!\!=\!\!& \frac{1}{\,64\pi^2 v_\phi^4\,}
\left\{M_h^4\pbrac{-\frac{3}{2}+\log\frac{M_h^2}{v_\phi^2}}
  + M_\X^4\pbrac{-\frac{3}{2}+\log\frac{M_\X^2}{v_\phi^2}} + 6M_W^4\pbrac{-\frac{5}{6}+\log\frac{M_W^2}{v_\phi^2}} \right.
\nonumber\\
&& \qquad\hspace*{7mm}  \left.
   + 3M_Z^4\pbrac{-\frac{5}{6}+\log\frac{M_Z^2}{v_\phi^2}}
   - 12M_t^4\pbrac{-1+\log\frac{M_t^2}{v_\phi^2}}
   - 6M_N^4\pbrac{-1+\log\frac{M_N^2}{v_\phi^2}}\right\}  ,
\\
\label{eq:B'}
\B^\prime &\!\!=\!\!& \frac{1}{\,64\pi^2 v_\phi^4\,}
    \pbrac{M_h^4 + M_\X^4 + 6M_W^4 +3M_Z^4 -12 M_t^4 - 6 M_N^4} \,.
\eeqa
\eeqs
In the coefficients above, we note that top quark carries a color factor $\,N_c^{}=3\,$,\,
while the three singlet neutrinos $\,\{\N_j^{}\}\,$ have an extra factor of
$\,1/2$\, due to their Majorana nature. We can readily verify that the
coefficients \eqref{ApBp} are related to the original definition \eqref{AB} via,
\begin{equation}
\label{coeffrel}
\A ~=~ \sin^4 \omega \pbrac{\A^\prime + \B^\prime \log \sin^2 \!\omega} \,,
\qquad~~
\B ~=~ \B^\prime \sin^4 \!\omega \ .
\hspace*{10mm}
\end{equation}
With \eqref{Lambda} and \eqref{ms}, we further deduce,
\beqs
\beqa
M_\sigma^2 &\!=\!& 8 \,\B^\prime v_{\phi}^2 \sin^2 \!\omega \ ,
\label{msfin}
\\[0mm]
\Lambda &\!=\!& v_\phi^{} \exp\tbrac{\frac{\A^\prime}{\,2\B^\prime\,}+\frac{1}{4}\,} ,
\label{Lambdafin}
\eeqa
\eeqs
where $\,\Lambda\,$ is the renormalization scale at which the flat direction
conditions of (\ref{mincond}) hold.
From \eqref{msfin}, the positivity condition of squared-mass $\,M_\sigma^2\,$ requires
$\,\B'>0$\,,\,  which takes the form
\beqa
\label{staboneloop}
M_\X^4 - 6 M_N^4 ~>~ 12M_t^4 - 6M_W^4 - 3M_Z^4 - M_h^4 \ .
\hspace*{10mm}
\eeqa
As anticipated, given the current data of mass measurements on the right-hand-side
of (\ref{staboneloop}),
this condition cannot be fulfilled by the SM particle content alone.
Hence, the Higgs sector of classically scale-invariant SM
has to be properly extended for realizing the radiative EWSB.
Finally, with (\ref{Lambdafin}) we can simplify the one-loop effective potential
\eqref{V1h0} by eliminating the coefficient $\,\A'\,$,
\beqa
\label{V1h-1}
V^{(1)}_{}(\phi) ~=~ \B' \phi^4 \(\log\frac{\phi^2}{v_\phi^2} \,-\,\hf\,\) \,.
\eeqa
We see that the one-loop potential \eqref{V1h-1} is bounded from below for
large values of $\,\varphi$\,,\, provided $\,\B' > 0\,$ which is ensured
by the positivity condition via \eqref{msfin} and realized in \eqref{staboneloop}.

Before concluding this section, let us summarize the present model.
Aside from the three exactly massless would-be Goldstone bosons
$\,(\pi^\pm_{},\,\pi^0_{})\,$ eaten by $(W^\pm,\,Z^0)$,\,
the scalar particle spectrum consists of the CP-even state $\,h\,$
and CP-odd state $\,\X\,$,\, with nonzero tree-level masses, and
an additional CP-even scalar $\sigma$,\, which
is a pseudo-Nambu-Goldstone boson of scale symmetry breaking,
with radiatively induced mass.
Furthermore, we have three singlet Majorana neutrinos $\,\N_j^{}\,$,\,
with masses generated from tree-level Yukawa interactions with singlet scalar \,$S$.\,
The Higgs potential \eqref{V0} includes six scalar-couplings, along with
two VEVs $\,(\vphi,\,\vet)\,$.\,  As explained, we can utilize the minimization condition
\eqref{mincond}  to express $\,\vet\,$ in terms of
$\,\vphi \simeq 246$\,GeV, and eliminate one of the coupling parameters in the
potential (say, $\lambda_\et^{}$)
according to the dimensional transmutation at scale $\Lambda$.\,
Furthermore, identifying the mass-eigenstate $\,h\,$ with the
LHC Higgs discovery $\,M_h^{} = 125$\,GeV,
we can eliminate one more coupling $\,\lambda_\phi^{}\,$ [as shown in
\eqref{masstreemincond}].  With these, we find that the present model
contains five input parameters in total, which, without losing  generality,
may be chosen as,
$\,\cbrac{\lambda_m^+,\, \lambda_m^-,\, \lambda_\X^{},\, \lambda_{\et\X}^{}, y_N^{}}$.\,
Given the defined mixing angle in \eqref{O}, and the tree-level masses of
$\X$ and singlet neutrinos $\,\N_j\,$ in \eqref{masstreemincond},
we can express the five independent inputs in terms of the more
physically transparent parameter set\,\footnote{Alternatively, it is possible to
choose $\,\cbrac{\sin\omega,\, M_\X^{},\, M_N^{},\, \lambda_\X^{},\,\lambda_{\eta\X}}$\,
as the inputs. But we find that the two sets of inputs are physically equivalent
for describing the parameter space.},
\beqa
\label{inputs}
\cbrac{\,\sin\omega,\, M_\X^{},\, M_N^{},\, \lambda_\X^{},\, \lambda_m^-\,}  \,.
\eeqa
The other couplings are non-independent and can be expressed as functions of them,
\begin{eqnarray}
&& \hspace*{-5mm}
\lambda_\phi^{}\,=\,\frac{\,3M_h^2\,}{v_{\phi}^2} \cos^2\!\omega \,,
\hspace*{8mm}
\lambda_{m}^+ \,=\, -\frac{\,M_h^2\,}{v_{\phi}^2} \sin^2\!\omega \,,
\hspace*{8mm}
\lambda_{\eta}^{} \,=\, \frac{\,3M_h^2\,}{v_{\phi}^2} \sin^2\!\omega\tan^2\!\omega \,,
\hspace*{8mm}
\nonumber\\
&& \hspace*{-5mm}
\lambda_{\eta\chi}^{} \,=\,
\(\frac{\,2M_\chi^2\,}{v_{\phi}^2}-\lambda_{m}^-\)\tan^2\!\omega \,,
\hspace*{8mm}
y_N^{} \,=\, \frac{\,M_N^{}\tan\omega\,}{\,\sqrt{2}\,v_\phi^{}\,} \,.
\label{eq:other-coup}
\end{eqnarray}
In the following section, we will systematically analyze the theoretical and
experimental constraints on the allowed parameter space of this model.


\vspace*{2mm}
\section{Experimental and Theoretical Constraints on the Parameter Space}
\label{bounds}
\vspace*{2mm}

In this section, we study various experimental and theoretical
constraints on the viable parameter space.
From experimental side, we will analyze the direct Higgs measurements
at the LHC and Tevatron, and the indirect electroweak precision tests.
For theoretical constraints, we will derive limits
from the perturbative unitarity, triviality and vacuum stability.
Finally, we present the combined numerical results and discuss
their physical implications.

\vspace*{2mm}
\subsection{{\bf Constraints from Direct Higgs Searches of the LHC}}
\vspace*{2mm}
\label{LHC-Higgs-bound}

As mentioned earlier, we will identify the CP-even Higgs boson $\,h$\, with the
125\,GeV new state discovered by the LHC.
Given the mixing between CP-even components of the
Higgs doublet $\,H\,$ and singlet $\,S\,$ in \eqref{O}, we note that $\,h\,$
couplings with other SM fields are suppressed by a factor of $\,\cos\omega\,$,
relative to the corresponding SM values.
We will perform a global fit of our model with the
LHC Higgs measurements, and derive the favored range of the mixing angle $\,\omega\,$.

To preform the global fit with LHC Higgs data, we start from a model-independent
effective Lagrangian formulation, where deviations of the associated couplings
from their SM values are taken as free parameters to be determined by data.
For the current analysis, our effective Lagrangian includes Higgs couplings to the
vector bosons, and heavy fermions (with top quark integrated out).
Thus, we can generally write down this effective Lagrangian,
\beqa
\label{Leff}
\mathcal{L}_{\text{eff}} &\!=\!&
   \(1\!+\delta_V^{}\) C^{\text{SM}}_{h WW} \,h\,W_{\mu}^{+}W^{-\mu}
  +\(1\!+\delta_V^{}\) C^{\text{SM}}_{h ZZ} \,h\,Z_{\mu}Z^{\mu}
  -\(1\!+\delta_b^{}\) C^{\text{SM}}_{h bb} \,h\,\bar{b}b
  -\(1\!+\delta_{\tau}^{}\) C^{\text{SM}}_{h\tau\tau} \,h\,\bar{\tau}\tau
\nonumber\\[2mm]
&&
  -\(1\!+\delta_c^{}\) C^{\text{SM}}_{h cc} \,h\,\bar{c}c
  +\(1\!+\delta_g^{}\) C^{\text{SM}}_{h gg} \,h\,G^a_{\mu\nu}G^{a\mu\nu}
  +\(1\!+\delta_{\gamma}^{}\) C^{\text{SM}}_{h\gamma\gamma}
  \,h\,A_{\mu\nu}A^{\mu\nu} \,,
\eeqa
where the coefficients $C^{\text{SM}}_{hXY}$ denote
the SM Higgs couplings to the fields $\,XY\,$,\, and potential
deviations are parametrized by the corresponding $\,\{\delta_{j}^{}\}$\,
which vanish in the pure SM.

For the present model, we find that the $h$ couplings in \eqref{Leff}
deviate from their SM values by the common suppression factor $\,\cos\omega$,\,
i.e., $\,\delta_{i} = \cos\omega - 1 = -\fr{1}{2} \omega^2 + O(\omega^4) < 0$\,.\,
With the LHC Higgs data, we can constrain the value of mixing angle $\,\omega\,$
since it is the only model parameter entering this analysis.
Also, the decay channel $\,h\to \sigma\sigma\,$
would be kinematically accessible for $\,M_h^{} > 2M_{\sigma}^{}\,$.\,
But, we find that the cubic $h$-$\sigma$-$\sigma$ coupling vanishes along
the flat direction \eqref{mincond} up to one-loop due to the nature of $\,\si\,$ being
pseudo-Goldstone of scale symmetry breaking.
Thus, the decay mode $\,h\to \sigma\sigma\,$ is absent.

Using the latest Higgs measurements from Lepton-Photon-2013\,\cite{LP2013}\cite{Tevatron},
we perform a global fit of the mixing parameter $\,\omega\,$ via effective
Lagrangian \eqref{Leff}, by minimizing the $\,\delta\chi^{2}\,$ function,
\beqa
\delta\chi^{2} ~=~ \sum_{ij}\,
(\hat{\mu}_{i}^{}-\hat{\mu}_{i}^{\text{exp}})
(\sigma^{2})^{-1}_{ij}
(\hat{\mu}_{j}^{}-\hat{\mu}_{j}^{\text{exp}})
\,,
\eeqa
where
$\,\hat{\mu}_{j}^{}=[\sigma\times\text{Br}]_j^{}/
   [\sigma\times\text{Br}]_j^{\text{sm}}$\,
denotes the Higgs signal strength for each given channel,
$\,j=\gamma\gamma,\,WW^*,\,ZZ^*,\,\bbbar,\,\tautaub\,$,\,
at ATLAS, CMS and Tevatron.  The error matrix is defined as,
\,$(\sigma^{2})_{ij}^{}=\sigma_{i}^{}\rho_{ij}^{}\sigma_{j}^{}$,\,
where $\,\sigma_i^{}$\, gives the corresponding error and
$\,\rho_{ij}^{}\,$ denotes the correlation matrix.
We present our findings in Table\,\ref{tab:hfit}.

\begin{table}[h]
\begin{center}
\begin{tabular}{c||c|c|c|c}
\hline\hline
&&&&
\\[-3mm]
$|\sin\omega|$ & 68\%\,C.L.\ & 95\%\,C.L. & Best Fit
& $\delta\chi^2_{\min}$/d.o.f.
\\
&&&&
\\[-3.5mm]
\hline
CMS  & $(0.14,~0.45)$ & $(0,~0.55)$ & $0.33$ & $0.36$
\\
\hline
All Data & $(0,~0.26)$ & $(0,~0.37)$ & -- & $0.85$
\\
\hline\hline
\end{tabular}
\caption{Global fit of Higgs mixing parameter $\,\sin\omega$\,
with the LHC and Tevatron data from Lepton-Photon-2013 \cite{LP2013}\cite{Tevatron}.}
\label{tab:hfit}
\end{center}
\end{table}

Table\,\ref{tab:hfit} shows that the CMS data alone prefers a nonzero Higgs mixing at
68\%\,C.L.\ and a best fit of $\,|\sin\omega|=0.33\,$,\,
while including all data from ATLAS/CMS and Tevatron puts a tighter
limit on the allowed range of mixing angle $\,\omega\,$,\, still consistent
with the SM case with zero mixing.
We note that the signal strengths measured by the current CMS data are lower than
the SM predictions in $\,h\to \gamma\gamma,WW,ZZ$\, channels \cite{LP2013}.
This is more consistent with the prediction from universal
$\cos\omega$ suppression in our present model, so our fit mildly prefers
$\,\omega\neq 0\,$ at 68\%\,C.L.\ (although still consistent with
$\,\omega = 0\,$ at 95\%\,C.L.).
Table\,\ref{tab:hfit} also shows that
the CMS data give a better fitting quality, as expected.
On the other hand, ATLAS gives somewhat enhanced signal strengths
in $\,h\to\gamma\gamma,ZZ$\, channels \cite{LP2013}.
Thus, the combined fit (including all data) allows less room for the $\cos\omega$ suppression,
and puts a tighter upper limit on $\,|\sin\omega|$\,.\, This combined fit is consistent with
$\,\omega = 0\,$ at 68\%\,C.L., but with a poorer fitting quality (due to the mild discrepancies
between the current CMS and ATLAS data mentioned above).

\vspace*{2mm}
\subsection{{\bf Constraints from Indirect Electroweak Precision Tests}}
\vspace*{2mm}
\label{ST-bound}

The present model contains two CP-even Higgs bosons in mass-eigenstates,
$\,h\,$ and $\,\sigma\,$.\,
The 125\,GeV SM-like Higgs boson $\,h\,$ has suppressed couplings relative to the
SM values by a factor of $\,\cos\omega$,\, whereas the couplings of
$\,\sigma\,$ scalar are proportional to the factor $\,\sin\omega$\,
[cf.\ \eqref{O}].  Thus, it is important to analyze the oblique corrections
via $S$ and $T$ parameters \cite{Peskin}. (It is easy to check that the contributions to
other oblique parameters are subleading as compared to $(S,\,T)$, and are
negligible for the current analysis.)  With electroweak precision tests
\cite{Baak:2012kk}, we can thus place indirect constraints on the parameter space.

Analytical expressions of the oblique corrections from an arbitrary number
of Higgs doublet and singlets were given in \cite{Grimus:2008nb}.
For our Higgs sector, we have the following results,
\beqs
\label{ST}
\beqa
\label{eq:dS}
\Delta S &\!=\!& \frac{\,\sin^2\!\omega\,}{24\pi}
\left[ \log R_{\sigma h} +
\hat G (M_\sigma^2,M_Z^2) - \hat G (M_h^2,M_Z^2) \right]  \,,
\\[1mm]
\label{eq:dT}
\Delta T &\!=\!& \frac{3\sin^2\!\omega}{16\pi\sin^{2}\!\theta_W^{} M_W^2}
\left[ M_Z^2 \(\frac{\log R_{Z\sigma}}{1-R_{Z\sigma}}
- \frac{\log R_{Zh}}{1-R_{Zh}}\)
- M_W^2 \( \frac{\log R_{W\sigma}}{1-R_{W\sigma}} - \frac{\log R_{Wh}}{1-R_{Wh}}\)
\right] \,,
\eeqa
\eeqs
where
\beqs
\label{eq:RGF}
\beqa
\label{eq:R}
R_{IJ}^{} &\!\!\!\equiv\!\!\!& {M_I^2}/{M_J^2} \,,
\\[1mm]
\label{eq:G}
\hat G_{IJ}^{} &\!\!\!\equiv\!\!\!&
-\fr{79}{3}+9R_{IJ}^{} -2R_{IJ}^2 + \pbrac{12 - 4R_{IJ} + R_{IJ}^2} \hat{F}_{IJ}^{}
+ \(-10+18R_{IJ} -6R_{IJ}^2+R_{IJ}^3
+ 9\frac{\,1\!+ R_{IJ}\,}{\,1\!- R_{IJ}\,}\)\log R_{IJ} \ ,
\hspace*{10mm}
\\[1mm]
\label{eq:F}
\hat F_{IJ}^{} &\!\!\!\equiv\!\!\!&
\left\{ \hspace*{-2mm}
\ba{ll}
\sqrt{R_{IJ}^{}(R_{IJ}^{}\!-4)\,}\,
\dis\log\!\fr{1}{2}\!\left|R_{IJ}^{}-2-\!\sqrt{R_{IJ}^{}(R_{IJ}^{}\!-4)\,}\right| \,,
 & \qquad (R_{IJ}^{}>4) \,,
\\[3.5mm]
2\sqrt{R_{IJ}^{}(4-R_{IJ}^{})\,} \, \arctan\!\sqrt{({4-R_{IJ}^{}})/R_{IJ}^{}\,} \,,
& \qquad (R_{IJ}^{}\leqq 4)\,.
\ea
\right.
\eeqa
\eeqs
In the formulas \eqref{ST}, $\theta_W^{}$ denotes the weak mixing angle, and
$\,M_\sigma^{}\,$ is the loop-induced mass of $\sigma$ scalar
[cf.\ Eq.\,\eqref{msfin}].   Given the experimental values of
$(M_W^{},\,M_Z^{})$ and the LHC data of $\,M_h^{} \simeq 125$\,GeV,\,
we find that the oblique parameters \eqref{ST} are functions
of $\,(\sin\omega,\,M_\sigma^{})\,$,\, where the radiative scalar mass $\,M_\sigma^{}\,$
still depends on $\,(M_\X^{},\,M_N^{})\,$,
as shown in Eqs.\,\eqref{msfin} and (\ref{eq:B'}).
Thus, imposing the electroweak precision data \cite{Baak:2012kk} and inputting
$\,M_N^{}\,$, we can use oblique corrections \eqref{ST} to
derive constraints on the parameter space of $\,(\sin\omega ,\,M_\X^{})\,$
or $\,(M_\si^{},\, M_\X^{})\,$,\,  as will be presented in Sec.\,\ref{results}.

\vspace*{2mm}
\subsection{{\bf Constraints from Perturbative Unitarity}}
\vspace*{2mm}
\label{uni-bound}

The SM has two essential features --- the perturbative renormalizablity and unitarity.
We require the same for the present extension.
It is trivial to say that the full $S$-matrix would be unitary
because computing an exact $S$-matrix cannot be practically done.
Therefore, it is important to study the
perturbative unitarity\,\cite{uni0}\cite{Lee:1977eg} of a given model,
which will ensure us to trust the theory predictions based on tree-level
and one-loop analyses.

For perturbative unitarity analysis, we are concerned with
the high energy behaviors of scattering amplitudes involving longitudinal
weak gauge bosons for the in/out states.
In the high energy regime, such scattering amplitudes can be given by
the corresponding Goldstone boson scattering amplitudes according to
the equivalence theorem \cite{ET}. Thus, we can perform a systematical
coupled-channel unitarity analysis for the scalar sector of our model.
Our Higgs sector contains three would-be Nambu-Goldstone bosons
$(\pi^\pm,\,\pi^0)$ eaten by $(W^\pm,\,Z^0)$, as well as
three neutral physical states $\,(h,\,\sigma,\,\X)$\,.\,

It is convenient to impose the unitarity condition on the partial wave amplitudes
of the scattering processes. The $s$-wave amplitude is inferred for a given $S$-matrix
element $\,\mathcal{T}\,$,
\beqa
\label{a0}
a_0 ~=~ \frac{1}{32\pi} \int_{-1}^1 \mathcal{T}\,\dd\cos\theta \ ,
\eeqa
and its unitarity condition is given by
$\,| \textrm{Re}\,a_0^{} | < \fr{1}{2}\,$ for the properly normalized in/out states.
For the coupled-channel analysis in the scalar sector,
we consider a set of electrically neutral in/out states,
\beqa
\label{eq:set-m}
\(
\left|\pi^+ \pi^-\right>,\,
\fr{1}{\sqrt{2}\,}\!\left| \pi^0\pi^0\right> ,\,
\fr{1}{\sqrt{2}\,}\!\left| h^0h^0\right> ,\,
\fr{1}{\sqrt{2}\,}\!\left|\sigma^0\sigma^0\right> ,\,
\fr{1}{\sqrt{2}\,}\!\left|\X^0\X^0 \right> ,\,
\left| \pi^0 h^0 \right> ,\,
\left| h^0\sigma^0 \right> ,\,
\left|\pi^0 \sigma^0\right> ,\,
\left|\pi^0\X^0\right> ,\,
\left|h^0\X^0\right> ,\,
\left|\sigma^0\X^0 \right>
\)  \,.
\eeqa
In the unitarity analysis, we mainly concern the high energy scattering
where the masses of in/out states are negligible. Thus, for convenience
we can work in their weak-eigenbasis, before the mass-diagonalization of
$(\phi,\,\eta)$ fields,
\beqa
\label{eq:set-0}
\(
\left|\pi^+ \pi^-\right>,\,
\fr{1}{\sqrt{2}\,}\!\left| \pi^0\pi^0\right> ,\,
\fr{1}{\sqrt{2}\,}\!\left| \phi^0\phi^0\right> ,\,
\fr{1}{\sqrt{2}\,}\!\left|\et^0\et^0\right> ,\,
\fr{1}{\sqrt{2}\,}\!\left|\X^0\X^0 \right> ,\,
\left| \pi^0\phi^0 \right> ,\,
\left| \phi^0\et^0 \right> ,\,
\left|\pi^0\et^0 \right> ,\,
\left|\pi^0\X^0  \right> ,\,
\left|\phi^0\X^0 \right> ,\,
\left|\et^0\X^0 \right>
\)  \,.
\eeqa
There are many tree-level diagrams contributing to the scattering processes
with the above in/out states.
They can be classified into three categories:
diagrams with quartic contact interactions,
diagrams with scalar-exchanges, and
diagrams with gauge boson exchanges.
For the high energy scattering regime,
the scalar-exchange contributions are suppressed by $\,E^{-2}\,$ relative to
the contact interactions. On the other hand, the diagrams with
gauge boson exchanges have momentum-dependent vertices, which may
compensate for the propagator-induced suppression and give $O(E^0)$ contributions.
But they are always proportional to the squared
electroweak gauge couplings which are subdominant
as compared with the pure quartic scalar couplings in the contact interactions.
Hence, similar to the SM case\,\cite{Lee:1977eg}, it suffices to consider the leading
contact interaction diagrams for the current unitarity analysis.

Using the weak-eigenbasis \eqref{eq:set-0} as in/out states,
we compute all possible two-body scattering processes with the
contact interactions \eqref{V0quart}, and derive the following matrix amplitudes,
%
\beqa
\label{eq:T}
&& \mathcal{T} \,=\,
\begin{pmatrix}
\mathcal{T}_1^{} & 0
\\[2mm]
0 & \mathcal{T}_2^{}
\end{pmatrix} ,
\hspace*{10mm}
\mathcal{T}_1^{} \,= -
\begin{pmatrix}
\fr{2}{3}\lambda_\phi^{} & \fr{1}{3\sqrt 2}\lambda_\phi^{}
& \fr{1}{3\sqrt 2}\lambda_\phi^{}
& \fr{1}{\sqrt 2}\lambda_m^+& \fr{1}{\sqrt 2}\lambda_m^-
\\[1.5mm]
\fr{1}{3\sqrt 2}\lambda_\phi^{} & \fr{1}{2}\lambda_\phi^{}
& \fr{1}{6}\lambda_\phi^{} & \fr{1}{2}\lambda_m^+ & \fr{1}{2}\lambda_m^-
\\[1.5mm]
\fr{1}{3\sqrt 2}\lambda_\phi^{} & \fr{1}{6}\lambda_\phi^{}
& \fr{1}{2}\lambda_\phi^{} & \fr{1}{2}\lambda_m^+
& \fr{1}{2}\lambda_m^-
\\[1.5mm]
\fr{1}{\sqrt 2}\lambda_m^+ & \fr{1}{2}\lambda_m^+
& \fr{1}{2}\lambda_m^+ & \fr{1}{2}\lambda_\et^{}
& \fr{1}{2}\lambda_{\et\X}^{}
\\[1.5mm]
\fr{1}{\sqrt 2}\lambda_m^- & \fr{1}{2}\lambda_m^-
& \fr{1}{2}\lambda_m^- & \fr{1}{2}\lambda_{\et\X}^{}
& \fr{1}{2}\lambda_{\X}^{}
\end{pmatrix} ,
\hspace*{16mm}
\nonumber
\\[2mm]
&& \mathcal{T}_2^{} \,=\, \textrm{diag}
  \(\fr{1}{3}\lambda_\phi^{},\,\lambda_m^+,\,\lambda_m^+,\,
    \lambda_m^-,\,\lambda_m^-,\,\lambda_{\et\X}^{}\) ,
\hspace*{5mm}
\eeqa
%
which has a coupled $5\times 5$ non-diagonal sub-block $\,\mathcal{T}_1^{}\,$.\,
Thus, substituting \eqref{eq:T} into \eqref{a0}, we derive the $s$-wave amplitude,
%
$\,a_0^{} = \mathcal{T}/(16\pi)\,$.\,
After diagonalization, we deduce the eigenvalues,
\beqa
\label{eq:a0-diag}
a_0^{}[\text{diagonal}] ~=~
-\frac{1}{\,16\pi\,} \,\text{diag}\(
\fr{1}{3}\lambda_\phi^{},~
\fr{1}{3}\lambda_\phi^{},~  x_1^{},~  x_2^{},~  x_3^{},~
\fr{1}{3}\lambda_\phi^{},~ \lambda_m^+,~
\lambda_m^+,~ \lambda_m^-,~ \lambda_m^-,~ \lambda_{\eta\X}^{}
\)
\eeqa
where $\,(x_1^{},\, x_2^{},\, x_3^{})\,$ are given by the roots
of the following cubic equation,
\beqa
\label{cubic}
&& 4 x^3 - 2\(2\lambda_\phi^{} +\lambda_\eta^{} +\lambda_\X^{}\) x^2
  + \tbrac{2\lambda_\phi^{}\lambda_\et^{} + 2\lambda_\phi^{}\lambda_\X^{}
   -4(\lambda_m^+)^2-4(\lambda_m^-)^2 - \lambda_{\et\X}^2 + \lambda_\et^{}\lambda_\X^{}}x
\hspace*{10mm}
\nonumber\\[1mm]
&& + \left[\lambda_\phi^{}\lambda_{\et\X}^2 - \lambda_\phi^{}\lambda_\et^{}\lambda_\X^{}
+ 2\lambda_\et^{} (\lambda_m^-)^2 + 2\lambda_\X^{} (\lambda_m^+)^2
- 4\lambda_{\et\X}^{} \lambda_m^+ \lambda_m^- \right] ~=~ 0 \,.
\eeqa
In the region of small Higgs mixing angle $\,\omega\ll 1\,$,\,
we find simple solutions for $\,(x_1^{},\, x_2^{},\, x_3^{})\,$ with
\beqa
x_{1,2}^{} ~\simeq\,
-\fr{1}{4}\left[\lambda_\chi\pm\!
\sqrt{16(\lambda_m^-)^2+4\lambda_{\eta\chi}^2+\lambda_\chi^2\,}\,\right] ,
\hspace*{10mm}
x_3^{}\simeq 0 \,.
\eeqa
For our numerical analysis in Sec.\,\ref{results}, we
will use the exact solutions of \eqref{cubic}.

In the above coupled channel analysis,
the eigenvalues in \eqref{eq:a0-diag} are all functions of our
input parameters \eqref{inputs}. Thus, imposing the unitarity condition
$\,|\text{Re} a_0^{} | < \fr{1}{2}\,$,\, we can derive constraints on the parameter space,
which will be presented in Sec.\,\ref{results}.

\vspace*{2mm}
\subsection{{\bf Constraints from Triviality and Vacuum Stability}}
\vspace*{2mm}
\label{TriStab-bound}

In this subsection, we analyze both triviality and stability bounds
for the present model.
The renormalization group (RG) equations
for SM gauge couplings $\,(g',\, g,\, g_s^{})$\,
and top Yukawa coupling \,$y_t^{}$\, are given by \cite{RG-SM},
\beqa
\label{eq:RGE-g}
&&
{\dd\, g'}/{\dd\, t} \,=\, (4\pi)^{-2}\(+\fr{41}{6}g^{\prime\,3}\) , ~~~\quad
{\dd\, g}/{\dd\, t}  \,=\, (4\pi)^{-2}\(-\fr{19}{6}g^3\) , ~~~\quad
{\dd\, g_s^{}}/{\dd\, t} \,=\, (4\pi)^{-2}\(-7g_s^3\) , \quad
\hspace*{10mm}
\nonumber
\\[1mm]
&&
{\dd\, y_t^{}}/{\dd\, t}  ~=~ (4\pi)^{-2} y_t^{}
\(\fr{9}{2}y_t^2 - 8g_s^2 -\fr{9}{4}g^2 - \fr{17}{12}g'^2\) ,
\eeqa
where $\,t=\log\mu\,$.\,
In addition, the Yukawa coupling \,$y_{N}^{}$\, of right-handed neutrinos
is defined in Eq.\,\eqref{LRHN}, and its RG equation reads, 
\beqa
\label{eq:RGE-yN}
{\dd\, y_N^{}}/{\dd\, t} ~=~ (4\pi)^{-2}\(\,+9y_N^3\,\) \,.
\eeqa
Thus, we can derive the RG evolutions for $\,(g',\, g,\,g_s^{},\,y_t^{},\,y_N^{})$\,.\,
The initial conditions for $\,(g',\, g,\,g_s^{})$ are defined at $\,\mu =M_Z^{}\,$,\,
while for $\,(y_t^{},\,y_N^{})$\,,\, we define,
$\,y_t^{}(M_t^{}) ~=~ {\sqrt{2}M_t^{}}/{v_\phi^{}}\,$ and
$\,y_{N}^{}(M_N^{}) ~=~ {M_{N}^{}}/(\!\sqrt{2}\,v_\eta^{})\,$.\,

It is also straightforward to compute the divergent parts of one-loop corrections
to the scalar quartic couplings in \eqref{couprel}. These include the vertex corrections
and wavefunction renormalizations. Thus, we derive their RG equations as follows,
\begin{equation}
\label{eq:RGE-QC}
\begin{split}
\dis{\dd\, \lambda_\phi^{}}/{\dd\, t}
~=~& \dis (4\pi)^{-2}
\left\{4\lambda_\phi^2+3\lambda_{m}^{+2}+3\lambda_{m}^{-2}
+3\lambda _\phi^{}\(4y_t^2-g'^2-3g^2\)
+\fr{9}{4}\tbrac{2g^4+\left(g^2+g'^2\right)^2-16y_t^4}\right\} \,,
\\[1mm]
\dis {\dd\, \lambda_\et^{}}/{\dd\, t}
~=~& \dis (4\pi)^{-2}
\left[ 3\lambda_{\eta}^{2}+12(\lambda_{m}^+)^{2}+3\lambda_{\eta\chi}^2
    + 24\lambda_{\eta}^{}y_{N}^{2} -288y_{N}^4\right] \,,
\\[1mm]
\dis {\dd\, \lambda_\X^{}}/{\dd\, t}
~=~& \dis (4\pi)^{-2}
\left[ 3\lambda_{\chi}^{2}+12(\lambda_{m}^-)^{2}+3\lambda_{\eta\chi}^2 \right] \,,
\\[1mm]
\dis {\dd\, \lambda_{m}^+}/{\dd\, t}
~=~& \dis (4\pi)^{-2}
\left\{ 4(\lambda_{m}^{+})^{2}+2\lambda_\phi^{}\lambda_{m}^+
  + \lambda_{\eta}^{}\lambda_{m}^+ +\lambda_{\eta\chi}^{}\lambda_{m}^-
  + \fr{3}{2}\lambda_{m}^+ \tbrac{4\(y_t^2 + 2y_N^2\) -g'^2-3g^2} \right\}  \,,
\\[1mm]
\dis {\dd\, \lambda_{m}^-}/{\dd\, t}
~=~& \dis (4\pi)^{-2}
\left\{ 4(\lambda_{m}^{-})^{2}+2\lambda_\phi^{}\lambda_{m}^- +\lambda_{\chi}^{}\lambda_{m}^-
        +\lambda_{\eta\chi}^{}\lambda_{m}^+
        +\fr{3}{2}\lambda_{m}^- \tbrac{4y_t^2 - g'^2 - 3g^2}\right\} \,,
\\[1mm]
\dis {\dd\, \lambda_{\et\X}^{}}/{\dd\, t}
~=~& \dis (4\pi)^{-2}
\left[ 4\lambda_{\eta\chi}^2+4\lambda_{m}^+\lambda_{m}^-
       +\lambda_{\eta\chi}^{}\(\lambda_{\eta}^{}+\lambda_{\chi}^{}\)
       + 12\lambda_{\eta\chi}^{}y_{N}^{2} \right] \,.
\end{split}
\end{equation}
The right-hand-sides of (\ref{eq:RGE-QC}) depend on all inputs
of the model parameters \eqref{inputs}.
Since the $\lambda_j^{}$'s are defined at a particular renormalization scale
$\,\mu=\Lambda\,$ where the tree-level flat direction conditions hold,
we will define the initial conditions of RG equations at $\,\Lambda\,$,\,
where $\,\Lambda\,$ is determined in terms of physical masses
$(M_\X^{},\, M_{N}^{})$\, via (\ref{Lambdafin}) and (\ref{ApBp}).

As shown in \eqref{eq:RGE-QC}, the scalar self-couplings $\{\lambda_j^{}\}$
have positive contributions to their beta functions
and thus tend to make them nonasymptotically free,
whereas the Yukawa couplings $(y_t^{},\,y_N^{})$ can give negative corrections
via box diagrams. When $\lambda_j^{}$'s dominate the beta functions,
these quartic scalar couplings will encounter Landau poles during the RG running.
Requiring $\{\lambda_j^{}\}$ not to diverge will thus impose constraints
(the triviality bounds) on the scalar masses\footnote{The pure SM with a 125\,GeV
Higgs boson is free from triviality up to Planck scale, since the SM triviality bound
only requires $\,M_h\lesssim 180\,$GeV \cite{SM-TB}.}\,
for a given UV cutoff $\,\cut\,$.\,
For practical numerical analysis, we will set a condition for all scalar couplings,
$\,\max\{\lambda_j^{}(\mu)\} < (4\pi)^2$\,,\, for $\,\mu < \cut\,$.\,
As we have explicitly checked, raising this upper limit from $\,(4\pi)^2\,$ up to infinity
does not produce any visible numerical difference.
Similar feature was also noted for the SM triviality analysis \cite{Sher}.

Then, we turn to the vacuum stability of the Higgs potential.
To ensure the stability of physical vacuum requires that the Higgs potential is bounded
from below. For the leading order, we employ the bounded-from-below conditions
(\ref{stabtree1})-(\ref{stabtree3}) for tree-level Higgs potential \eqref{V0quart},
with couplings improved by one-loop RG runnings \eqref{eq:RGE-QC}
to ensure vacuum stability at high scales.
This will constrain the parameter space for each given cutoff scale $\,\cut$.\,
Besides, the one-loop Higgs potential \eqref{V1h-1} is stabilized
under the condition \eqref{staboneloop}.
For the present analysis we study the conditions for a stable physical vacuum.
As an alternative, it may be possible that the vacuum is merely meta-stable
\cite{SM-TB,Sher}, which would be worth of a future study.

\vspace*{2mm}
\subsection{{\bf Viable Parameter Space: Combined Numerical Aanlysis}}
\label{results}
\vspace*{2mm}

\begin{figure}
\begin{center}
\hspace*{-4mm}
\includegraphics[width=0.46\textwidth]{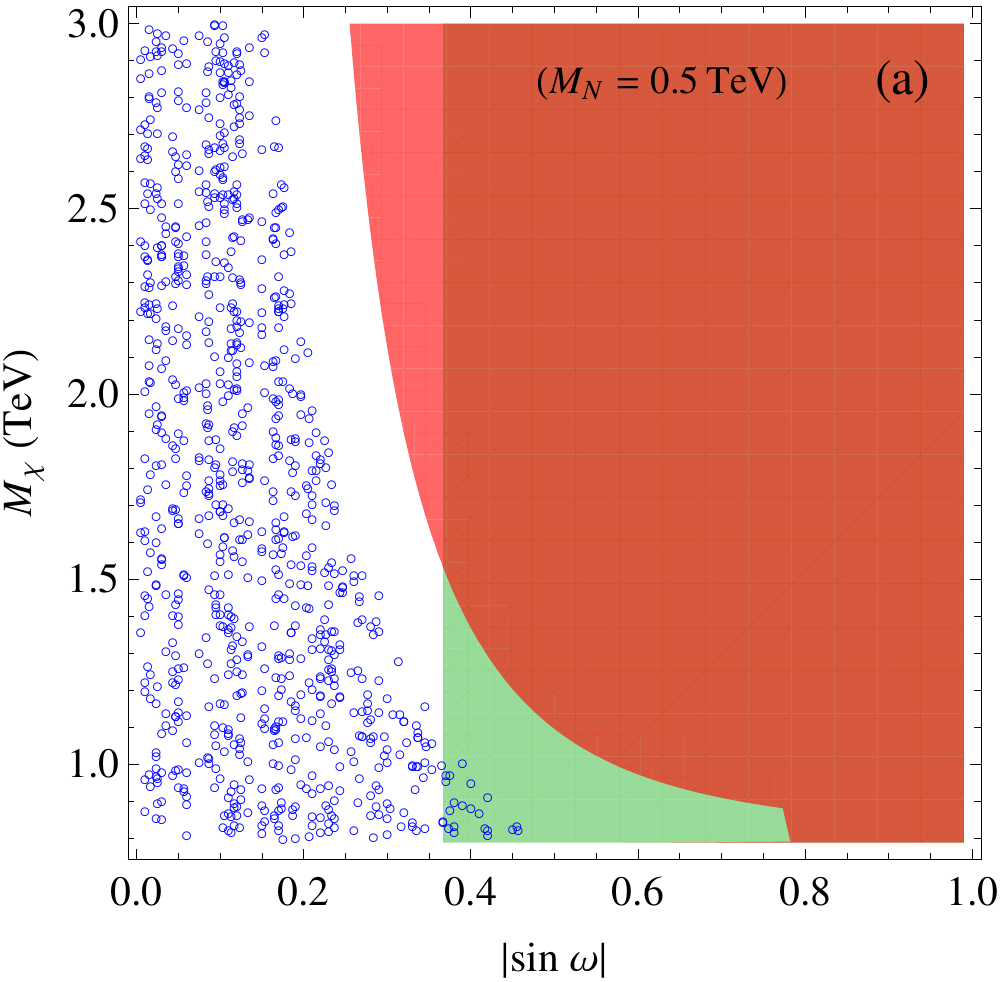}
\hspace*{2mm}
\includegraphics[width=0.46\textwidth]{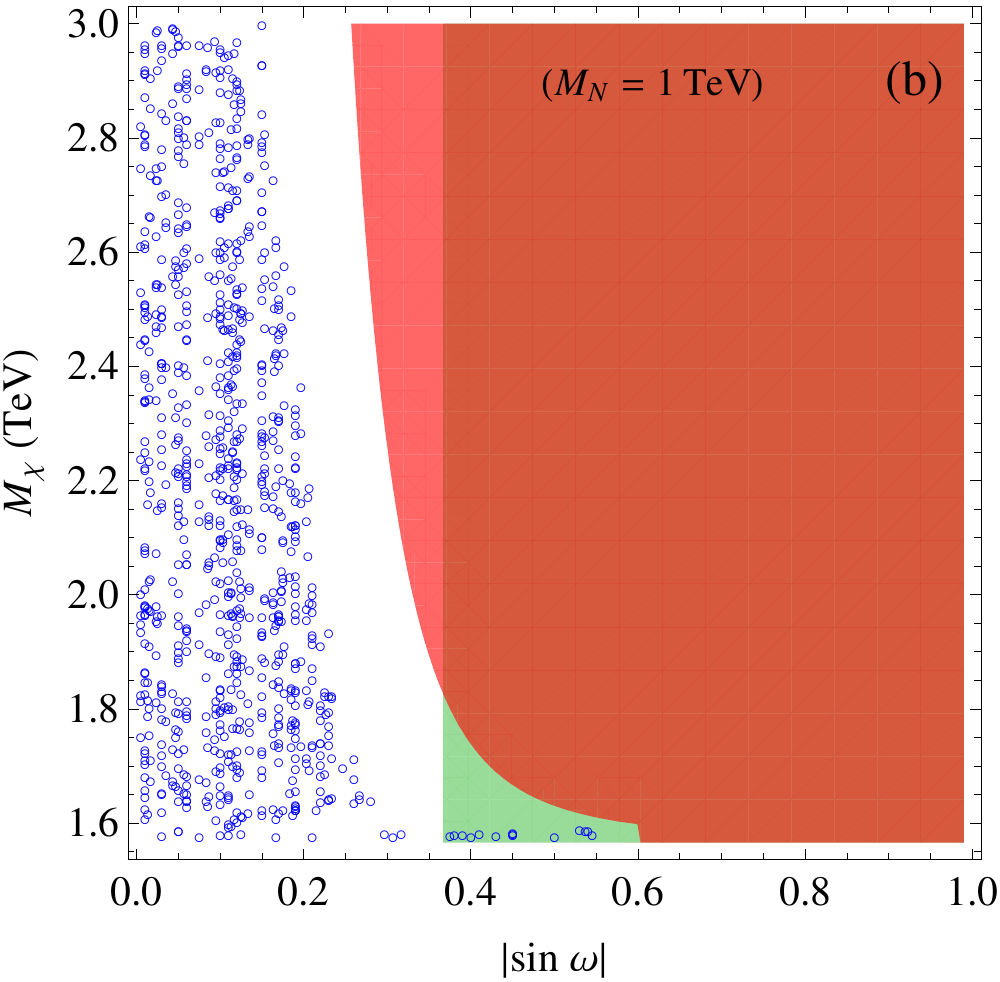}
\\[3mm]
\hspace*{-4mm}
\includegraphics[width=0.46\textwidth]{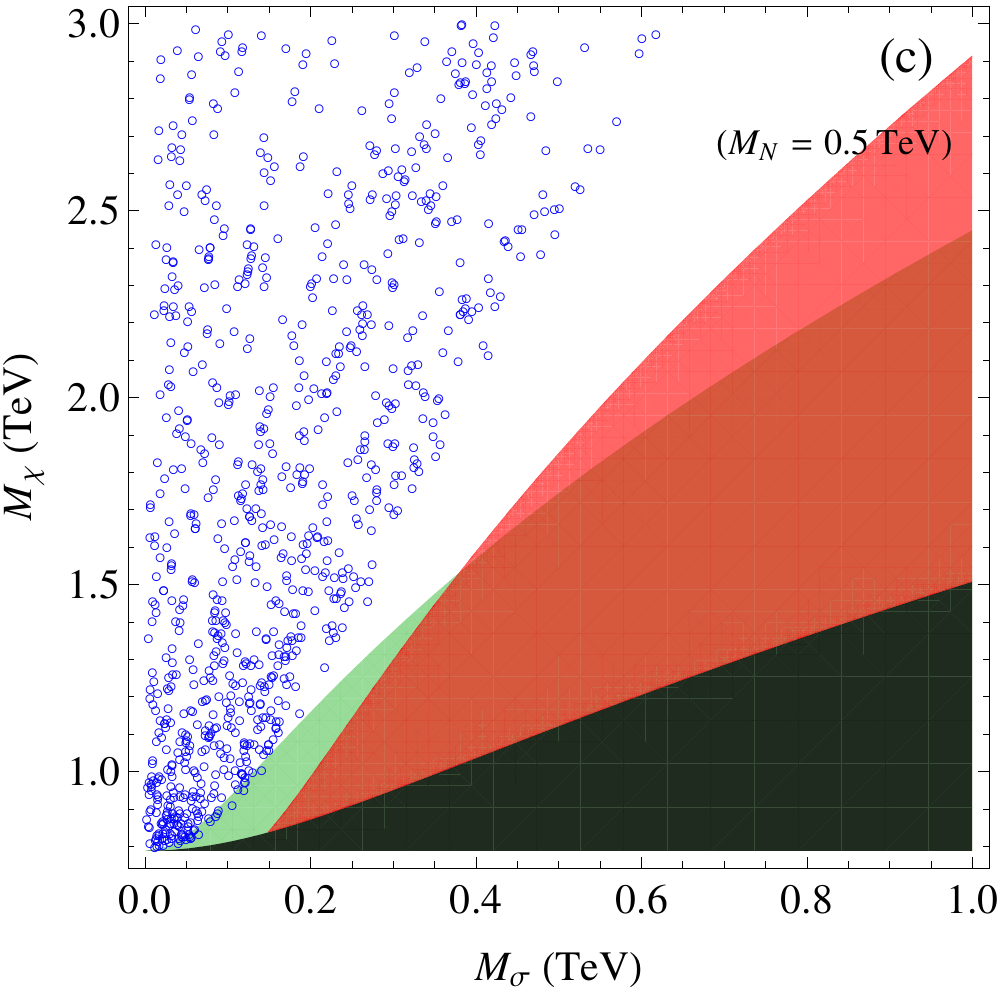}
\hspace*{2mm}
\includegraphics[width=0.46\textwidth]{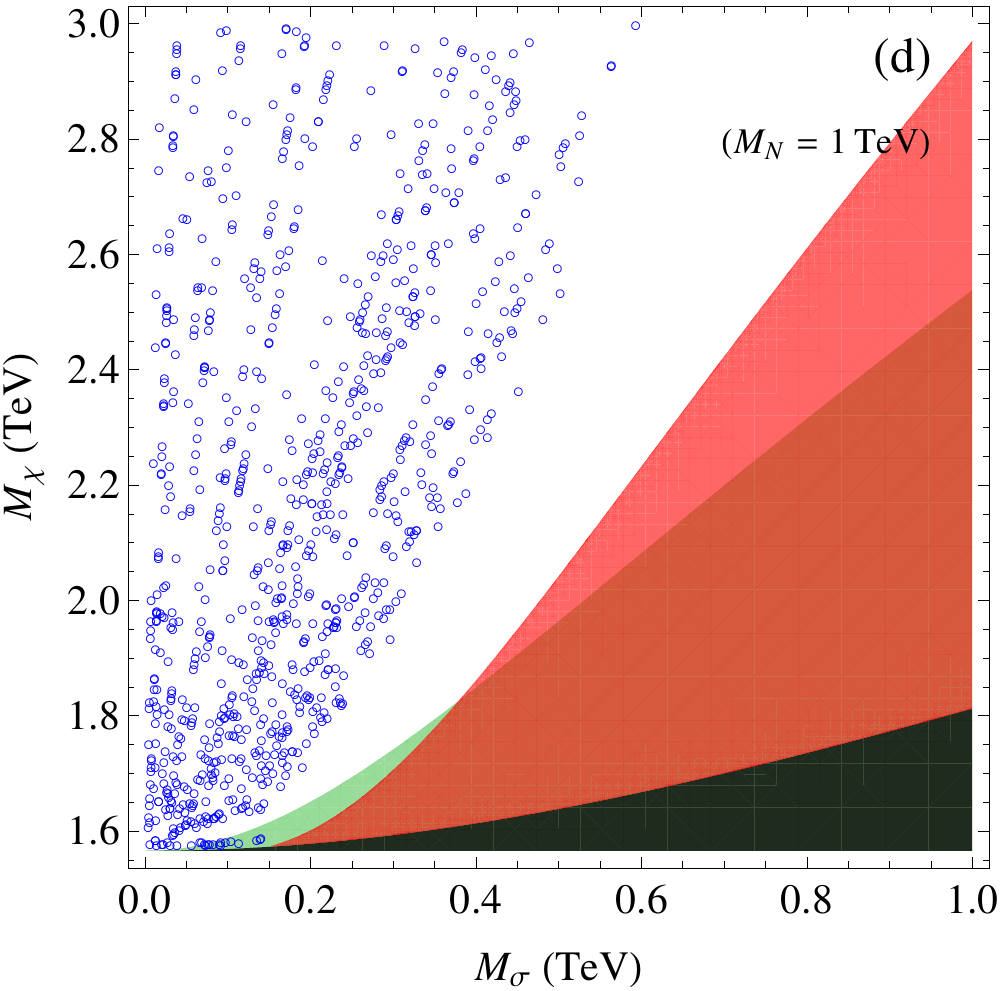}
\end{center}
\vspace*{-4mm}
\caption{Experimental and theoretical constraints on the parameter space of
$\,|\sin\omega|-M_\X^{}\,$ in plots (a)-(b), and
of $\,M_\si^{}-M_\X^{}\,$ in plots (c)-(d).
In each plot, the red region is excluded by the electroweak precision tests at 95\% C.L.,
while the green region is excluded (95\%\,C.L.) by the global fit of direct Higgs
measurements at the LHC and Tevatron (as partially overlapped by the red contour).
In plots (c)-(d), the shaded black region is forbidden due to
$\,|\sin\omega|\leqq 1\,$.\,
The scattered blue-dots are simulated in each plot to represent the
parameter space, allowed by the triviality, stability and perturbative unitarity bounds.
We set two benchmark values of right-handed neutrino masses,
$\,M_N^{} = 0.5, 1$\,TeV, for the plots (a)(c) and (b)(d), respectively.
}
\label{fig:1}
\end{figure}

In this subsection, we systematically present numerical analysis
of the viable parameter space by imposing the
experimental and theoretical constraints studied
in Sec.\,\ref{LHC-Higgs-bound}-\ref{TriStab-bound}.

Inspecting the five independent input parameters of \eqref{inputs},
we will analyze the viable parameter space in the two-dimensional plane
of $\,\{\sin\omega,\,M_\X^{}\}\,$ or $\,\{M_\sigma^{},\, M_\chi^{}\}\,$
by scanning the allowed ranges of
scalar couplings $\,\{\lam_\X^{},\,\lam_m^-\}\,$
for each given sample mass $\,M_N^{}\,$ of right-handed neutrinos.

For the experimental constraints in Sec.\,\ref{LHC-Higgs-bound}-\ref{ST-bound},
we note that the LHC Higgs measurements of $\,h\,(125\,\text{GeV})\,$
only depend the Higgs mixing angle ($\sin\omega$), while the
electroweak oblique corrections \eqref{ST} are functions of
$\{\sin\omega,\,M_\sigma^{}\}$,\, or, equivalently,
$\{\sin\omega,\,M_\X^{},\, M_N^{}\}$  via \eqref{msfin}.
For the numerical analysis, we will set two benchmark values
of the right-handed neutrino mass, $\,M_N^{} =0.5,\,1$~TeV.\,
In Fig.\,\ref{fig:1}(a)-(b), we first present the experimental constraints
in the shaded regions.  The red region in each plot is excluded
by the precision tests at 95\%\,C.L.\  via
oblique corrections  \eqref{eq:dS}-\eqref{eq:dT}.
The vertical green band displays the excluded parameter region on
$\,|\sin\omega|\,$ by the global fit of direct Higgs measurements
of the LHC and Tevatrion at 95\%\,C.L. (Table\,\ref{tab:hfit}),
which is partly overlapped by the red contour of precision bound.
In Fig.\,\ref{fig:1}(c)-(d), we impose the same experimental constraints
in the plane of $\,M_\si^{} - M_\X\,$.

For the theoretical constraints in Sec.\,\ref{uni-bound}-\ref{TriStab-bound},
they depend on all input parameters of \eqref{inputs}.
For Fig.\,\ref{fig:1}(a)-(b),
we simulate 900 random points in $\,|\sin\omega| - M_\X^{}\,$ plane
for the remaining two input couplings $\,\{\lam_\X^{},\,\lam_m^-\}\,$
under the unitarity bound, the triviality bound, and stability conditions.
These scattered points in Fig.\,\ref{fig:1} are
statistically represented by the small blue-dots.
Our unitarity constraints are derived from tree-level partial wave analysis,
while the triviality and vacuum stability bounds invoke loop corrections and
RG runnings up to the UV cutoff scale $\,\cut$\,.\,
In the numerical simulations, we scan over the range of
$\,\cut \geqq 10^5\,$GeV for RG evolutions.  Note that
the positivity condition \eqref{staboneloop} also sets a generic lower bound on
$\,M_\X^{}\,$  for a given input of $\,M_N\,$.\,
We find, $\,M_\X^{}> 0.79\,(1.57)\,$TeV for $\,M_N=0.5\,(1.0)\,$TeV,\,
as indicated in Fig.\,\ref{fig:1}.
It is evident that the full analysis favors relatively small mixing
between the Higgs doublet and singlet with
$\,|\sin\omega|\lesssim 0.3\,$.\,
In parallel, for Fig.\,\ref{fig:1}(c)-(d), we simulate 900 random points
in $\,M_\si^{} - M_\X\,$ plane for input couplings $\,\{\lam_\X^{},\,\lam_m^-\}\,$
under the same theoretical constraints as in plots (a)-(b),
shown by the small blue-dots.
We see that for each given $\,M_\si^{}\,$,\, the mass of $\,\X\,$ receives a lower bound,
while for a given $\,M_\X^{}\,$,\, the $\,\sigma\,$ mass acquires an upper bound.
For instance, taking $\,M_\X^{} \leqq 2\,$TeV will impose an upper limit
$\,M_\si^{}\lesssim 400\,(300)$\,GeV for $\,M_N^{}=0.5\,(1.0)$\,TeV.\,
Our parameter space predicts $\,M_\si^{}\,$ to be significantly lighter than $\,M_\X^{}\,$.\,
This is expected, since the mass of pseudo-Nambu-Goldstone boson $\,\si\,$
is radiatively generated.

\vspace*{1mm}

As a final remark, we note that the definition of the Higgs mixing
angle $\,\omega\,$ flips sign if we identify the 125\,GeV state as the
pseudo-Nambu-Goldstone boson $\,\sigma\,$ of the scale symmetry breaking,
where the small mixing angle $\,\omega\,$ will
correspond to a small singlet VEV $\,v_\et^{}$\,.\,
In this case, the theoretical bounds heavily constrain the small mixing
region, while the large mixing range remains excluded by
the experimental bounds. After detailed numerical scan of parameter space,
we conclude that this inverse-identification scenario is ruled out
by the combined theoretical and experimental constraints of Sec.\,\ref{bounds}.
Besides, we find that replacing the complex singlet $\,S\,$ by a real singlet
scalar is also excluded by these constraints. Thus, the present model gives the
minimal viable construction.

\vspace*{2mm}
\section{Conclusions and Discussions}
\label{conclusion}
\vspace*{2mm}

The recent LHC Higgs discovery \cite{LHCnew}\cite{LP2013} opens up a new era
for particle physics, pressing us to seek natural Higgs mechanism and explore
the associated new physics (including the dark matter candidate).

\vspace*{1mm}

In this work, we constructed the minimal viable extension of the SM
with classical scale symmetry. It realizes radiative electroweak symmetry breaking (EWSB)
\`{a} la Coleman-Weinberg and gives a natural solution to the hierarchy problem.
In addition to a SM-like light Higgs boson $\,h\,$ of mass 125\,GeV,
it predicts two new states at weak scale:
one CP-even Higgs $\,\si\,$ serving as the pseudo-Nambu-Goldstone boson
of scale symmetry breaking, and a CP-odd scalar singlet $\,\X\,$ providing
the dark matter (DM) candidate.
Furthermore, the model nicely accommodates three right-handed Majorana
neutrinos $\,\N_{1,2,3}^{}\,$ and generates light neutrino masses via TeV scale seesaw.

\vspace*{1mm}

In Sec.\,\ref{Sec:2.1}-\ref{Sec:2.2}, we presented the model,
whose Higgs sector contains the SM Higgs doublet plus a new complex singlet.
We formulated the scale-invariant and CP-symmetric Higgs potential (\ref{V0})
at tree-level, and determined the flat direction (\ref{mincond}),
as well as realizing the TeV scale seesaw of light neutrino masses (\ref{eq:seesaw})
via Yukawa interaction (\ref{LRHN}).
In Sec.\,\ref{Sec:2.3}, we systematically studied the radiative EWSB
\`{a} la Coleman-Weinberg, and derived complete Higgs mass-spectrum in
(\ref{masstreemincond}) and (\ref{msfin}).

\vspace*{1mm}

In Sec.\,\ref{LHC-Higgs-bound}, we first analyzed experimental constraints
from global fit of the direct Higgs measurements at the LHC and Tevatron
(cf.\ Table\,\ref{tab:hfit}).
Then, in Sec.\,\ref{ST-bound}, we derived oblique corrections in (\ref{ST})
and analyzed the corresponding electroweak precision tests.
In Sec.\,\ref{uni-bound}-\ref{TriStab-bound}, we studied theoretical constraints
from unitarity, triviality and vacuum stability for this model.
Combining both the experimental and theoretical constraints, we analyzed
the viable parameter space in Sec.\,\ref{results}, which are
presented in Fig.\,\ref{fig:1}(a)-(d).

\vspace*{1mm}

Finally, we discuss signals of the predicted new
$\,\si\,$ and $\,\X\,$ bosons at the upcoming runs of the LHC\,(14\,TeV).
The $\,\si\,$ boson has couplings to $WW/ZZ$ and quarks/leptons
under the suppression\footnote{This suppression is due to the Higgs
mixing \eqref{O} for $\,(h,\,\si)\,$.\, So, the production and decay signals of $\,\si\,$
are similar to that of the heavier Higgs state $\,H^0\,$ from new physics models with
extended two-Higgs-doublet sector \cite{exd-H}.}
of $\,\sin\omega\,$.\,
Hence, it can be produced at the LHC via gluon fusions $\,gg\to \si\,$,\,
with $\,\si\to ZZ,WW\,$ as its major detection channels,
where $ZZ$ and $WW$ decay leptonically.
It also has an interesting decay mode, $\,\sigma\to hh\,$,\, for
$\,M_\si^{}> 250\,$GeV.\,
On the other hand, the DM candidate $\,\X\,$ can be produced
in pair at the LHC, giving raise to missing energy. Since $\,\X\,$ only appears in the
Higgs potential \eqref{V0quart} and couples to $\,h\,$ and $\,\si\,$
besides its self-coupling, we expect its major LHC production channel
comes from the associate production,
$\,pp\to Zh^* \to Z\X\X\,$ and  $\,pp\to Wh^* \to W\X\X\,$,\,
with $\,Z\to \ell^-\ell^+$ and $\,W\to \ell\nu\,$ ($\ell = e,\mu$).

\vspace*{1mm}

We also note that our model predicts a relatively heavy scalar DM particle $\,\X\,$
with mass of $\,O(\text{TeV})\,$.\,
The positivity condition \eqref{staboneloop} generally sets a lower bound on $\,M_\X\,$,\,
and requires, $\,M_\X^{}> 0.79\,(1.57)\,$TeV for $\,M_N=0.5\,(1.0)\,$TeV.\,
We verify its viability as a cold DM by computing the thermal relic density.
Since $\,M_\X^{}\,$ is heavier than all other particles in the model,
it is reasonable to consider that all particles are in thermal equilibrium around
the time when $\,\chi\,$ freezes out. There is considerable rate for
a pair of $\,\chi\,$ annihilating into other two-body final states,
which arise from contact interactions and exchanges of scalars.
Thus, given the measured DM relic density $\,\Omega_{\text{DM}}^{}\,$,\,
we can derive constraints on the parameter space \eqref{inputs}.
As for the DM direct detection, the relevant couplings concern the
\,$\X$-$\X$-$h$\, or \,$\X$-$\X$-$\si$\, vertices with $\,h\,$ or $\,\si\,$
interacting with the SM fermions. The corresponding effective contact
interactions of this DM with nucleons can be tested by direct detection experiments,
such as XENON100 \cite{XENON100}, and CDMS-II \& EDELWEISS \cite{CDMSII}.
So far, the XENON100 detection gives the best bound on spin-independent cross sections
of TeV scale DM \cite{XENON100}. A systematical DM analysis for our present model
is beyond the current scope and will be given elsewhere.

\vspace*{5mm}
\noindent
{\bf Acknowledgments}
\\[1.5mm]
We are grateful to Estia Eichten, Robert Foot, Bob Holdom, Francesco Sannino,
Kimmo Tuominen, and Raymond Volkas for valuable discussions.
This work was supported by National NSF of China (under grants
11275101, 11135003) and National Basic Research Program (under grant
2010CB833000). A.F.\ was supported in part by Tsinghua Outstanding
Postdoctoral Fellowship.


\end{document}